\documentclass[journal]{IEEEtran} 

\IEEEoverridecommandlockouts

\usepackage{graphicx}
\usepackage{color}

\usepackage{cite}
\usepackage{amsmath}
\usepackage[caption=false]{subfig}

\usepackage{amssymb}

\usepackage[acronym]{glossaries}

\usepackage[noabbrev]{cleveref}

\Crefname{equation}{}{}

\newtheorem{proposition}{\bf Proposition}

\newcommand{\buyer}{Buck}
\newcommand{\seller}{Seth}

\newcommand{\SIM}{simulation scenario}
\newcommand{\MB}{model-based}
\newcommand{\DD}{data-driven}

\newcommand{\wbl}{Weibull}

\newcommand{\losDuration}{\tau}
\newcommand{\coherenceTime}{\losDuration_0}

\newcommand{\confidence}{\gamma}

\newcommand{\outage}{\epsilon}

\newcommand{\rxpower}{R}
\newcommand{\rxpowerThreshold}{\rxpower_0}

\newcommand{\distributionScale}{\sigma}
\newcommand{\distributionShape}{\xi}
\newcommand{\distributionCombined}{\vect{\theta}}
\newcommand{\distribution}{F}
\newcommand{\distributionEst}{\distribution_{\distributionCombined}}

\newcommand{\estimated}[1]{\hat{#1}}

\newcommand{\sample}{m}
\newcommand{\SAMPLE}{M}
\newcommand{\sampleSet}{\set{\SAMPLE}}

\newcommand{\positiveSet}{\sampleSet^{+}}
\newcommand{\negativeSet}{\sampleSet^{-}}

\newcommand{\lossFunction}{L}

\newcommand{\failureRate}{Z}

\newcommand{\order}{n}

\newcommand*{\myfigfactor}{0.7}
\newcommand*{\myfigfactorx}{0.305}

\newcommand{\optMinimize}{\min}
\newcommand{\optMaximize}{\max} 
\newcommand{\subjectTo}{\text{subject to}}

\newcommand{\expect}{\mathbb{E}\,}
\newcommand{\probability}{\text{Pr}}

\newcommand{\naturalset}{\mathbb{N}}

\newcommand{\vect}{\boldsymbol}

\newcommand{\set}[1]{\mathcal{#1}}

\newcommand{\optimal}{^\star}
\newcommand{\inverse}{^{-1}}

\usepackage{forloop}
\newcounter{loopcntr}
\newcommand{\rpt}[2][1]{%
	\forloop{loopcntr}{0}{\value{loopcntr}<#1}{#2}%
}

\newcommand{\subgroup}[1]%
{\rlap{\smash{%
	\newcount\cnt%
	\cnt \numexpr#1\relax%
	\advance\cnt -1\relax%
	$\tabcolsep=.1em\begin{tabular}[t]{|l}\multicolumn{1}{l}{}\\%
	\rpt[\cnt]{\\}
	\\\hline\end{tabular}$%
}}}

\newcounter{myRefCount}

\definecolor{shadecolor}{gray}{0.95}
\newcounter{myQuestion}
{
	\begin{shaded*}
	\stepcounter{myQuestion}
	\enumerate[\setlength{\labelwidth}{-.5em}]
	\item[\textbf{\themyQuestion.}]
	\itshape
}
{
	\endenumerate
	\end{shaded*}%
}
\makeatother

\newacronym{tx}{Tx}{transmitter}

\newacronym{rx}{Rx}{receiver}

\newacronym{mle}{MLE}{maximum likelihood estimation}

\newacronym{mmw}{mmWave}{millimeter wave}

\newacronym{los}{LOS}{line-of-sight}

\newacronym{rss}{RSS}{received signal strength}

\newacronym{pdf}{PDF}{probability distribution function}

\newacronym{cdf}{CDF}{cumulative density function}

\newacronym{ccdf}{CCDF}{complementary cumulative density function}

\newacronym{gpr}{GPR}{Gaussian process regression}

\newacronym{nn}{NN}{neural network}

\newacronym{relu}{ReLU}{rectified linear unit}

\newacronym{satlins}{SATLILNS}{symmetric saturating linear}

\newacronym{urllc}{URLLC}{ultra reliable low-latency communication}

\newacronym{urc}{URC}{ultra-reliable communication}

\newacronym{lr}{LR}{linear regression}

\newacronym{mse}{MSE}{mean square error}

\newacronym{mlp}{MLP}{multilayer perceptron}

\newacronym{ml}{ML}{machine learning}

\newacronym{tp}{TP}{true positive}

\newacronym{fp}{FP}{false positive}

\newacronym{fn}{FN}{false negative}
\allowdisplaybreaks

\begin{document}

\title{%
	Predictive Ultra-Reliable Communication: A Survival Analysis Perspective
}

\author{
\IEEEauthorblockN{
	Sumudu Samarakoon, \IEEEmembership{Member IEEE}, 
	Mehdi Bennis, \IEEEmembership{Fellow IEEE}, \\
	Walid Saad, \IEEEmembership{Fellow IEEE} 
	and 
	M\'{e}rouane Debbah, \IEEEmembership{Fellow IEEE}
	\\}
\vspace{-30pt}
\thanks{This work is supported by Academy of Finland 6G Flagship (grant no. 318927) and project SMARTER, projects EU-ICT IntellIoT and EU-CHISTERA LearningEdge, Infotech-NOOR and NEGEIN, and the U.S. National Science Foundation under Grant CNS-1836802.}
\thanks{Sumudu Samarakoon and Mehdi Bennis are with University of Oulu, Finland (e-mail: \{sumudu.samarakoon,mehdi.bennis\}@oulu.fi).}
\thanks{Walid Saad is with ireless@VT, Bradley Department of Electrical and Computer Engineering, Virginia Tech, Blacksburg, VA (e-mail: walids@vt.edu).}
\thanks{M\'{e}rouane Debbah is with Universit\'{e} Paris-Saclay, CNRS, CentraleSup\'{e}lec, 91190, Gif-sur-Yvette, France (e-mail: merouane.debbah@centralesupelec.fr) and the Lagrange Mathematical and Computing Research Center, 75007, Paris, France.}
}

\maketitle
\nopagebreak[4]

\begin{abstract}

\Gls{urc} is a key enabler for supporting immersive and mission-critical 5G applications.
Meeting the strict reliability requirements of these applications is challenging due to the absence of accurate statistical models tailored to \gls{urc} systems.
In this letter, the wireless connectivity over dynamic channels is characterized via statistical learning methods.
In particular, \emph{model-based} and \emph{data-driven} learning approaches are proposed to estimate the non-blocking connectivity statistics  over a set of training samples with no knowledge on the dynamic channel statistics.
Using principles of \emph{survival analysis}, 
the reliability of wireless connectivity is measured in terms of the probability of channel blocking events.
Moreover,
the maximum transmission duration for a given reliable non-blocking connectivity is predicted in conjunction with 
the confidence of the inferred transmission duration. 
Results show that the accuracy of detecting channel blocking events is higher using the model-based method for low to moderate reliability targets requiring low sample complexity. 
In contrast, the data-driven method shows higher detection accuracy for higher reliability targets at the cost of $100\times$ sample complexity.

\end{abstract}

\glsresetall
\section{Introduction}\label{sec:introduction}

Next-generation wireless services, such as mission and safety critical applications, require \gls{urc} that provision certain level of communication services with guaranteed high reliability \cite{Saad2019,Park2020}.
Realizing this in the absence of statistical models tailored to tail-centric \gls{urc} systems is known to be a daunting task  \cite{Popovski2014,Bennis2018}.

Towards enabling \gls{urc}, the majority of the existing literature relies on system-level simulations-based brute-force approaches leveraging packet aggregation and spatial, frequency, and temporal diversity techniques
 \cite{Bennis2018,Chaccour2020}
while some 
assume perfect or simplified/approximated models of the system (i.e., stationary channel and traffic models) \cite{Hur2016}.
However, such approximations may fail to characterize the tail statistic accurately, and thus, may inadequate to fulfill the reliability targets of \gls{urc} \cite{Angjelichinoski2019}.
	In this view, \gls{ml} techniques have been used in the context of \gls{urc} including low-latency aspects with a focus on channel modeling and prediction
\cite{Nishio2019,Kasgari2019,She2020,Hou2018}.
These works are mostly data-driven and assume the availability of large amounts of data.
All prior works focusing on channel modeling can be used to optimize transmission parameters preventing communication outages in terms of loss of \gls{rss} due to channel blockage.
Here, a channel blocking event represents a period during which the \gls{rss} remains below a predefined target threshold
and the channel transitions from non-blocking to blocking events are analogous to the so-called \emph{survival time} \cite{3gpp2018}.
Characterizing such channel transitions is useful to determine highly reliable transmission intervals under the absence of knowledge of channel statistics, which has not been done in the existing literature.

The transitions between non-blocking and blocking can be cast as lifetime events (birth-to-death) of the channels.
Analyzing the time to an event (e.g., a channel transition) and rate of event occurrence are the prime focuses of \emph{survival analysis}  \cite{Abernethy1983}.
The applications of survival analysis span a multitude of disciplines including medicine (life expectancy and mortality rate from a disease), 
engineering (reliability of a design/component),
economics (dynamics of earnings and expenses),
and finance (financial distress analysis) \cite{Li2013,Matz2002,Laitinen2005}.
Therein, either model-based or model-free methods can be adopted.
Hence, we adopt the analogy behind survival analysis to investigate non-blocking connectivity over wireless links.

The main contribution of this work is to characterize the statistics of non-blocking connectivity durations under the absence of knowledge on the dynamic wireless channel statistics.
In this view, we consider a simplified communication setting consists of a single \gls{tx}-\gls{rx} pair communicating over dynamic channels with a fixed transmission power in order to characterize the transmission duration guaranteeing a reliable non-blocking connectivity.
The underlying challenge with the above analysis lies in assuming or acquiring the full knowledge of non-blocking duration statistics, which is unfeasible.
Hence, we address two fundamental questions: 
i) how to accurately model the non-blocking duration statistics without the knowledge of channel statistics?
and
ii) how to characterize the confidence bounds for reliable transmission durations inferred from the devised non-blocking duration statistics?
To this end, we consider an exemplary scenario of a buyer named \buyer{} who plans to purchase radio resources for a \gls{urc} system from a seller named \seller{}.
Here, \buyer{} needs to evaluate the radio resources in terms of the transmission periods guaranteeing low blocking probabilities under different connectivity durations and the statistics of transmission periods to enable \gls{urc}.
For this purpose, \seller{} wishes to reliably evaluate the connectivity failure statistics, i.e., via \emph{survival analysis}, using a set of non-blocking connected duration samples $\sampleSet$ over dynamic channels.
However, \seller{} must address key questions related to the training data set $\sampleSet$:
\emph{i)} does it contain sufficient samples?
\emph{ii)} how confident am I with the reliability measures obtained using $\sampleSet$?
	and
\emph{iii)} is it beneficial to improve the prediction confidence by investing in additional sampling?
Towards addressing these questions, we first cast the problem of finding the maximum transmission duration yielding a predefined low blockage probability as an optimization problem.
Therein, we adopt a tractable parametric representation for the probabilistic model of channel failures.
To estimate the parameters, a minimization of a loss function that captures the gap between the true-yet-unknown channel failure probability and the parametric representation is formulated.
To minimize the aforementioned loss function, we adopt two approaches: a \emph{model-based approach} that assumes a known prior probabilistic model following \wbl{} survival analysis, and a \emph{data-driven approach} that uses function regression via \glspl{nn}.
For both techniques, wireless connectivity is analyzed in terms of the conditional failure statistics, namely the statistics of the time to fail under given connectivity durations, and their confidence bounds followed by an evaluation based on simulations. 

\section{System Model and Problem Formulation}\label{sec:system_model}

\begin{figure}[t!]
	\centering
	\includegraphics[width=.9\linewidth]{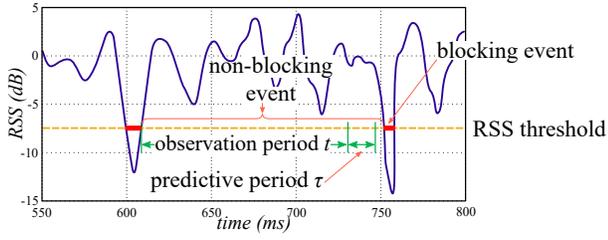}
	\caption{An illustration of the channel blocking and non-blocking durations for a given \gls{rss} threshold.}
	\label{fig:los_illustrate}
\end{figure}

Consider a one-way communication system in which a \gls{tx} sends data to a \gls{rx} over a correlated flat fading channel.
Due to channel and mobility dynamics, the \gls{rss} at the \gls{rx} fluctuates over time.
For a given target \gls{rss} $\rxpowerThreshold$, we define the non-blocking connectivity probability (also called \emph{survival probability}) as $\probability(\rxpower_t \geq\rxpowerThreshold)$ where $\rxpower_t$ represents the \gls{rss} over the duration $[0,t]$.
In \gls{urc}, the goal is to identify a predictive period $\losDuration > 0$ that guarantees a low \emph{conditional} blocking probability after observing a non-blocking connectivity over a duration of $t$, i.e., $\probability( \rxpower_{\losDuration+t} < \rxpowerThreshold | \rxpower_{t} \geq \rxpowerThreshold ) \leq \outage$ given an outage probability $\outage$, as illustrated in Fig. \ref{fig:los_illustrate}.

In this considered system, neither the channel dynamics nor the statistics of non-blocking connectivity are known a priori.
Our objective is to obtain a reliable measure of the \gls{cdf} of the blocking events (or the \gls{ccdf} of the connected durations), i.e., $\distribution(t) = \probability(\rxpower_t \leq \rxpowerThreshold)$.
Once $\distribution(t)$ is characterized, the conditional failure probability at an observation period $t$ will be:
\begin{equation}\label{eqn:conditional_prob}
	\failureRate_t(\losDuration) = 
	\probability( \rxpower_{\losDuration+t} < \rxpowerThreshold | \rxpower_{t} \geq \rxpowerThreshold )
	\textstyle = \frac{\distribution(t+\losDuration) - \distribution(t)}{1 - \distribution(t)}.
\end{equation} 
Then, determining the transmission duration followed by the observation period of $t$ for a given target reliability $1-\outage$, is formulated as follows:
\begin{equation}\label{eqn:optimization_observed_period}
{\optMaximize}_{} \quad  \losDuration, \qquad
\subjectTo \quad \failureRate_t(\losDuration) \leq \outage.
\end{equation}
For a known and analytically tractable $\distribution(\cdot)$, 
the solution of \eqref{eqn:optimization_observed_period} is given by
$\losDuration\optimal = \distribution\inverse \big( \outage + (1-\outage)\distribution(t) \big) - t$.
However, $\distribution(\cdot)$ is unknown 
due to the absence of channel statistics and the lack of accurate modeling of time-varying system parameters (e.g., network geometry, mobility, scattering coefficients, etc.),
and thus, needs to be estimated.

\section{Estimating $\distribution(\cdot)$}\label{sec:statistical_model}

To estimate the non-blocking duration distribution, a parametric representation of the \gls{cdf} $\distributionEst(\cdot)$ with parameter vector $\distributionCombined$ can be adopted. 
Here, $\distributionCombined$ is calculated using a set $\sampleSet$ of $\SAMPLE$ connected duration samples.
For this purpose, a loss function $\lossFunction(\cdot)$ that captures the gap between the estimated and actual \glspl{cdf} needs to be minimized over the sample set $\sampleSet$ as follows:
\begin{equation}\label{eqn:optimization_problem}
{\optMinimize}_{\distributionCombined} \quad
\lossFunction_{\sampleSet}(\distributionEst,\distribution).
\end{equation}
Towards solving \eqref{eqn:optimization_problem}, we consider two approaches:
i) \emph{model-based approach:} assuming a known prior probabilistic model to derive the distribution parameters $\distributionCombined$ corresponding to the prior distribution using \eqref{eqn:optimization_problem}
and
ii) \emph{data-driven approach:} using \gls{nn}-based function regression over $\sampleSet$ where $\distributionCombined$ is the \gls{nn} model to be learned from the data.

\subsection{Model-Based Approach}\label{sec:model_based}

The events of non-blocking durations can be interpreted as the lifetimes of connected periods that are terminated by the drop of \gls{rss} below a target threshold, which then is followed by blocking events.
In this view, the statistical tools of survival analysis are suitable for characterizing the non-blocking connectivity durations.
In particular, \wbl{} distribution is the most widely used lifetime data model due to its relation to various families of distributions (uniform, exponential, Rayleigh, generalized extreme value, etc.)
\cite{Abernethy1983}.
Accordingly, the non-blocking connectivity durations can be modeled by a \wbl{} distribution,
\begin{equation}\label{eqn:weibull_cdf}
\distributionEst(t) = 
1 - 
e^{-(t/\distributionScale)^{\distributionShape}},
\end{equation}
where $\distributionCombined=(\distributionScale,\distributionShape)$ is parameterized by the scale ($\distributionScale$) and shape ($\distributionShape$) parameters.
To find the most likely parameter values that fit \eqref{eqn:weibull_cdf} to $\sampleSet$, we use \gls{mle}.
In this regard, we define the loss function $\lossFunction_{\sampleSet}(\distributionCombined) = - \sum_{\sample} \log f_{\distributionCombined}(t_\sample)$ where  $f_{\distributionCombined}(t) = \frac{\distributionShape}{\distributionScale} \big( \frac{t}{\distributionScale} \big)^{\distributionShape-1} e^{-(t/\distributionScale)^\distributionShape}$ is the \wbl{} \gls{pdf}.
Due to the non-convex nature of the objective function, the estimated parameters $\estimated{\distributionCombined}$ can be found using numerical methods (e.g., stochastic gradient decent).
Using $\estimated{\distributionCombined}$, the failure probability in \eqref{eqn:conditional_prob} becomes:
\begin{equation}\label{eqn:fail_rate_wbl}
	\failureRate_t(\losDuration,\estimated{\distributionCombined})
	= \textstyle 1 - 
	\exp \big( ({t}/{\estimated{\distributionScale}})^{\estimated{\distributionShape}} - 
		({(t+\losDuration)}/{\estimated{\distributionScale}})^{\estimated{\distributionShape}}
		\big).
\end{equation}
Then, the solution for \eqref{eqn:optimization_observed_period} will be:
\begin{equation}\label{eqn:optimal_observed_duration}
	\losDuration\optimal = \estimated{\distributionScale} \big( 
	\textstyle ({t}/{\estimated{\distributionScale}})^{\estimated{\distributionShape}} 
	- \ln(1 - \outage) \big)^{1/\estimated{\distributionShape}} - t.
\end{equation}

Note that the reliable transmission duration $\losDuration\optimal$ hinges on the training data set $\sampleSet$.
Therefore, it is important to provide the margins of confidence for the derived values.
To evaluate the confidence bounds, we adopt the likelihood ratio bounds method \cite{Verrill2007} given as:
\begin{equation}\label{eqn:confidence_MLE}
2 \big( \lossFunction_{\sampleSet}(\distributionCombined) - \lossFunction_{\sampleSet}(\estimated{\distributionCombined}) \big) \geq
\chi^2_{\confidence,\SAMPLE},
\end{equation} 
where $\chi^2_{\confidence,\SAMPLE}$ are the Chi-squared statistics with probability $\confidence$ and degree-of-freedom $\SAMPLE$, and $\distributionCombined$ is the unknown true parameter, respectively.
For example, $\confidence=0.95$ yields 95\% confidence interval of the parameter estimation.
Since we are interested in evaluating the confidence for $\losDuration\optimal$ rather than $\distributionCombined=(\distributionScale,\distributionShape)$, we first find 
$\distributionScale = \sqrt[\distributionShape]{ ( t^\distributionShape - (t+\losDuration)^\distributionShape ) / \ln (1-\outage) }$
using \eqref{eqn:optimal_observed_duration} and, then, \eqref{eqn:confidence_MLE} can be modified as follows:
\begin{equation}\label{eqn:confidence_MLE_modified}
\textstyle
\lossFunction_{\sampleSet}(\sqrt[\distributionShape]{ ( t^\distributionShape - (t+\losDuration)^\distributionShape ) / \ln(1-\outage) },\distributionShape) - \lossFunction_{\sampleSet}(\estimated{\distributionCombined}) =
\frac{\chi^2_{\confidence,\SAMPLE}}{2}.
\end{equation} 
Note that a closed-form expression cannot be derived for \eqref{eqn:confidence_MLE_modified}
which calls for numerical solutions
(e.g., trust-region algorithm \cite{Nocedal2006}).
Since both $\losDuration$ and $\distributionShape$ are unknown in \eqref{eqn:confidence_MLE_modified}, for some $\delta>0$, several priors for $\distributionShape$ from $[\estimated{\distributionShape}-\delta,\estimated{\distributionShape}+\delta]$ are selected first.
By solving \eqref{eqn:confidence_MLE_modified} for each of the above choices, a set of solutions $\{\losDuration\}$ is obtained, from which the confidence bounds of $\losDuration\optimal$ are calculated.
In addition to $\losDuration\optimal$, its mean and variance can be analytically derived using \eqref{eqn:fail_rate_wbl}.
\begin{proposition}\label{thm:wbl_moments}
	The $N$th moment of the non-blocking connectivity duration $t+\losDuration$ under the observation duration $t$ is:
	\begin{equation}\label{eqn:nth_moment}
		\expect [(t+\losDuration)^N] = 
		\distributionScale^N e^{(t/\distributionScale)^\distributionShape}
		\Gamma\big( (t/\distributionScale)^\distributionShape; 1 + N/\distributionShape \big),
	\end{equation}
	where $\Gamma(\alpha,\beta) = \int_{\alpha}^{\infty} x^{\beta-1} e^{-x} dx$ is the \emph{upper incomplete gamma} function.
\end{proposition}
\begin{IEEEproof}
	See Appendix \ref{appndx:wbl_moments}.
\end{IEEEproof}
Using the above result, the mean and variance of the remaining connectivity durations at time $t$ can be obtained from $\expect [t+\losDuration] - t$ and $\expect [(t+\losDuration)^2] - \expect^2 [t+\losDuration]$, respectively.

\subsection{Data-Driven Approach}\label{sec:data_driven}

The main drawback of the model-based approach is its susceptibility to model drift whereby the statistics of the actual observations may differ from the \wbl{} model.
Hence, estimating $\distributionEst(\cdot)$ by using the empirical distribution of samples $\sampleSet$ is preferable.
Next, a data-driven approach based on a \gls{nn}-based regression is presented.

First, a subset of data samples $\sampleSet_t = \{t_\sample | t_\sample \geq t, t_\sample\in\sampleSet \}$ is collected 
for a given observation period $t$.
Then, the empirical distribution of the non-blocking duration samples in $\sampleSet_t$ is numerically evaluated so that a set of labeled training data tuples $\{(t_\sample, s_\sample)\}$ are generated.
Here, $s_\sample$ is the \gls{cdf} value of $t_\sample$ calculated using the empirical distribution, which yields the corresponding failure distribution.
The loss function is the \gls{mse} between the true and estimated failure probabilistic values, i.e., 
$\lossFunction_{\sampleSet_t}(\distributionCombined) = \frac{1}{\SAMPLE_t} \sum_\sample \big( s_\sample - \failureRate_t(t_\sample,\distributionCombined) \big)^2$
 where $\failureRate_t(\cdot,\distributionCombined)$ is modeled using a \gls{mlp} with model parameters $\distributionCombined$.
To solve \eqref{eqn:optimization_problem}, \gls{mlp} uses $(t_\sample, t_\sample^2, \dots, t_\sample^\order)$ up to an order of $\order$ (to avoid under-fitting) as the input, $s_\sample$ as the output, and the \gls{mse} loss $\lossFunction_{\sampleSet_t}(\distributionCombined)$ as the empirical loss function.
By training the \gls{mlp} in a supervised manner, $\failureRate_t(\cdot,\distributionCombined)$ is derived.
Finally, $\losDuration\optimal$ that satisfies $\failureRate_t=\outage$ is obtained.
Note that the accuracy of $\failureRate_t(\cdot,\distributionCombined)$ relies on 
i) both quality and quantity of $\sampleSet_t$,
ii) the model complexity of $\distributionCombined$, and
iii) choice of the input size $\order$.

The $N$th raw moment of the remaining non-blocking connectivity for an observation duration $t$ will be:
\begin{equation}\label{eqn:nn_statistics}
	\textstyle \expect[\losDuration^N|t] = \int_{0}^{\infty} 
	\losDuration^{N-1} \big( 1 - \failureRate_t(\losDuration,\distributionCombined) \big) d\tau.
\end{equation}
First, the conditional probabilities are calculated from the trained \gls{nn} model over a sequence of $\losDuration = \delta k$ remaining connectivity durations with $k\in\naturalset$ and small $\delta>0$.
Then, approximating the integrations in \eqref{eqn:nn_statistics} to numerical summations, the first and second moments of the remaining connectivity durations can be obtained.

\section{Simulation Results and Analysis}\label{sec:results}

Here, we evaluate the characterization of non-blocking statistics obtained via the proposed \MB{} and \DD{} methods.
For our simulations, we consider a time correlated Rayleigh flat fading channel model defined in \cite{Rappaport1996}.
While we define a slotted time-based transmission with a slot duration of $\coherenceTime = 1\,\text{ms}$, for improved measurement accuracy, we consider a sampling frequency of $4\,\text{kHz}$.
For training, up to $10,000$ non-blocking connectivity duration samples are collected and for testing, additional $30,000$ samples are used.
Here, an \gls{rss} threshold of $\rxpowerThreshold = -8\,$dB is used for a unit transmit power.
For the data-driven approach, we use an \gls{mlp} with two fully connected hidden layers with sizes of ten and six
and \gls{relu} activations. 
The output layer of the \gls{mlp} is a single node with a symmetric saturated linear transfer function.

\begin{figure}[!t]
	\centering
	\subfloat[Model-based estimation.]{
		\includegraphics[width= .9\linewidth]{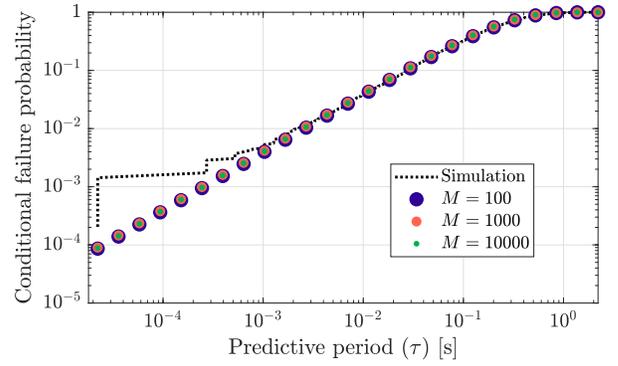}
		\label{fig:sampleComplexityModel}
	}
	
	\subfloat[Data-driven estimation for different order of input sizes $\order\in\{1,10\}$.]{
		\includegraphics[width= .9\linewidth]{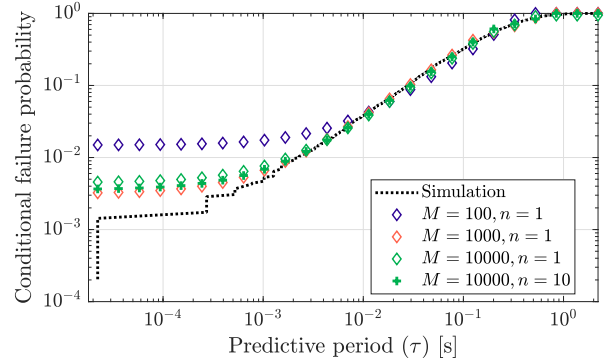}
		\label{fig:sampleComplexityData}
	}
	\caption{Comparison of the conditional failure probability estimation at $t = 0.3$\,s for different sample complexities $\SAMPLE\in\{100, 1000, 10000\}$.}
	\label{fig:curver_fitting}
\end{figure}

Fig. \ref{fig:curver_fitting} compare the conditional failure probability regression performance of both the \MB{} and the \DD{} approaches over the simulated data referred to as ``\SIM'' for different sample complexities, i.e., various choices of training sample sizes $\SAMPLE\in\{100, 1000, 10\,000\}$.
From Fig. \ref{fig:sampleComplexityModel}, we observe that the \MB{} design is almost invariant over the choices of sample complexities due to the accurate fit over probabilities above $10^{-2}$.
As the probability decreases, the simulation results will deviate from the trend of higher probabilities.
However, the \MB{} method, which relies on the prior \wbl{} model, fails to capture this deviation.
In contrast, the \DD{} regression is susceptible to the lack of training samples as illustrated in Fig. \ref{fig:sampleComplexityData}.
Moreover, it can learn the trends using data samples and thus, the \DD{} approach learns the low-probability behavior of the \SIM{} as well.
In addition, Fig. \ref{fig:sampleComplexityData} shows that increasing the order $n$ from one to ten slightly improves the regression. 
This improvement is due to the fact that we consider the input as a tenth order polynomial of the connectivity duration instead of order one.

\begin{figure}[!t]
	\centering
	\includegraphics[width= \myfigfactor\linewidth]{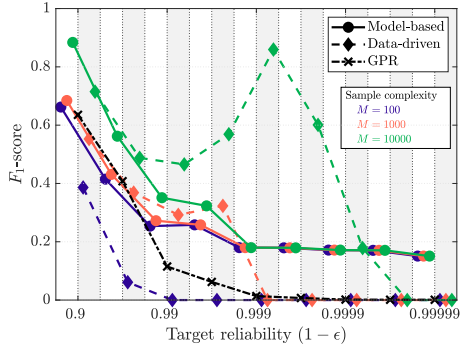}
	\caption{Detection of blocked events based on the predicted duration $\losDuration\optimal$ at $t=0.3$\,s.}
	\label{fig:fscore}
\end{figure}

Fig. \ref{fig:fscore} compares the detection of channel blocking events based on the predicted duration $\losDuration\optimal$ from \MB{} and \DD{} methods in terms of \emph{F-score}:
$F_1 = \frac{ \sum \text{TP}}{\sum\text{TP}+(\sum\text{FP}+\sum\text{FN})/2}$
based on the events of \emph{\gls{tp}}, \emph{\gls{fp}}, and \emph{\gls{fn}}
\cite{Blair1979}.
We first empirically partition the test connectivity durations dataset $\sampleSet'$ into two groups for a given reliability target $(1-\outage)$: 
the \emph{positive} group $\positiveSet_{\outage}$ consisting of the smallest $\outage$ fraction of non-blocking durations and
the rest composes the \emph{negative} group $\negativeSet_{\outage}$.
With this partitioning, for any test sample $\sample\in\sampleSet'$ there are three observation categories:
i) \textbf{\gls{tp}}: if $\sample < \losDuration\optimal$ and $\sample\in\positiveSet_{\outage}$,
ii) \textbf{\gls{fp}}: if $\sample < \losDuration\optimal$ but $\sample\in\negativeSet_{\outage}$,
and
iii) \textbf{\gls{fn}}: if $\sample \geq \losDuration\optimal$ with $\sample\in\positiveSet_{\outage}$.
In addition, for the purpose of comparison, a \gls{gpr}-based channel estimation method proposed in \cite{Karaca2012} is adopted to predict consecutive non-blocking durations, which is referred to as the ``GPR'' baseline.
Fig. \ref{fig:fscore} shows that as the sample complexity increases, the uncertainty of the estimated $\losDuration\optimal$ decreases and blocked events are accurately detected, achieving higher $F_1$.
For large $\outage$, the estimated $\losDuration\optimal$ from the \MB{} approach can accurately detect the channel blocking events (i.e., the lower tail) yielding high $F_1$.
As $\outage$ decreases, the \MB{} method based on the \wbl{} distribution bias deviates from the actual data distribution even if the increasing training sample size $\SAMPLE$ increases. 
From this result, we observe that the accuracy of channel blocking detection degrades by factors of $2\times$ to $3\times$ as shown in Fig. \ref{fig:fscore}.
In contrast, the \DD{} approach characterizes the lower tail better than the \MB{} method when a sufficiently large number of  training data is available.
For a small $\SAMPLE$, the detection accuracy of the \DD{} method approaches to zero with decreasing $\outage$, because of the lack of training data in the positive set $\positiveSet_{\outage}$ of the size of $\outage\SAMPLE$.
Hence, increasing $\SAMPLE=100$ to 
$1000$ and then to 
$10\,000$ improves the blocked event detection accuracy from $F_1=0$ to $0.32$ and $0.49$ at $\outage=10^{-2}$ and from $F_1=0$ to $0.82$ at $\outage=10^{-3}$, respectively, highlighting the importance of the sample complexity in \DD{} methods.
The GPR baseline outperforms both proposed methods with $\SAMPLE\in\{100,1000\}$ only for small reliability targets $\outage\geq 0.05$.
Due to the uncertainty in \gls{gpr}, 
higher prediction errors can be observed for tighter reliability targets, 
resulting in a low $F_1$.

\begin{figure}[!t]
	\centering
	\includegraphics[width= .9\linewidth]{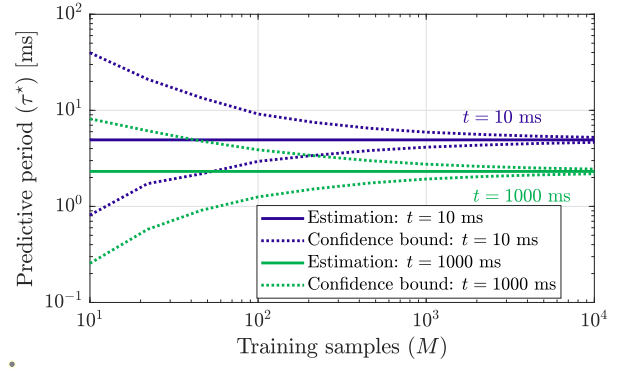}
	\caption{Confidence limits of predictive transmission durations obtained using the \MB{} approach for observed durations $t\in\{10, 1000\}$\,ms.}
	\label{fig:confidence}
\end{figure}

Fig. \ref{fig:confidence} illustrates the impact of sample complexity on the confidence bounds of the predicted transmission durations at $t\in\{0.01, 1\}$\,s derived using the \MB{} approach.
Here, a 95\% confidence interval (i.e., $\confidence=0.95$) is used.
From Fig. \ref{fig:confidence}, we can see that \gls{mle} with few samples yields large uncertainty in $\losDuration\optimal$ while the uncertainty decays as $\SAMPLE$ increases due to the monotonic decreasing nature of $\chi^2_{\confidence,\SAMPLE}$ with $\SAMPLE$. 
This underscores the tradeoff between the model parameter uncertainty and the cost of data collection.

\begin{figure*}
	\centering
	\subfloat[Impact of transmit power on the predicted transmission duration.]{
		\includegraphics[width= \myfigfactorx\linewidth]{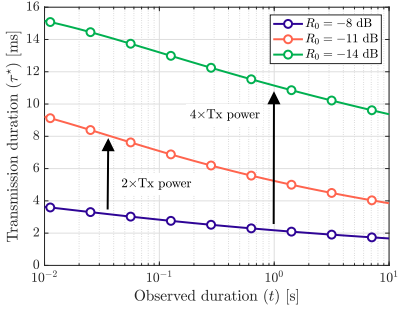}
		\label{fig:hazard}
	}
	\hspace{.005\linewidth}
	\subfloat[Expected time to fail for the observed data and predictions.]{
		\includegraphics[width= \myfigfactorx\linewidth]{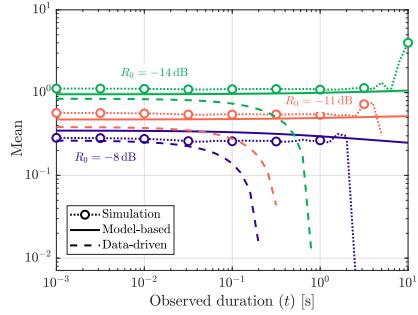}
		\label{fig:meanChange}
	} 
	\hspace{.005\linewidth}
	\subfloat[Variance of failure time for the observed data and predictions.]{
		\includegraphics[width= \myfigfactorx\linewidth]{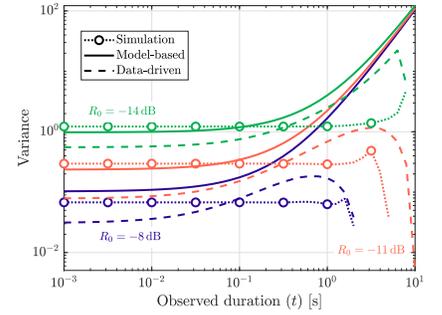}
		\label{fig:varChange}
	}
	\caption{Impact of transmit power on the predicted duration ($\losDuration\optimal$) ensuring $(1-\outage)$ reliability (left), expected time to fail (middle), and its variance (right).}
	\label{fig:powerImpact}
\end{figure*}

The impact of the transmit power is investigated in Fig. \ref{fig:powerImpact}.
Since $\rxpowerThreshold = -8\,$dB is used with a unit transmit power, a $2\times$ and $4\times$ increase in transmit power are captured with $\rxpowerThreshold$ of $-11\,$dB and $-14\,$dB, respectively.  
The effects of increasing transmit power on the predicted connectivity durations derived from the \MB{} approach are presented in Fig. \ref{fig:hazard}.
Clearly, the non-blocking connectivity can be significantly enhanced via increased transmission power.

For a given observation duration $t$, the mean and variance of the remaining non-blocking connectivity durations over the simulated data and the estimations based on both the \MB{} and the \DD{} methods are shown in Figs. \ref{fig:meanChange} and \ref{fig:varChange}, respectively.
Note that the \SIM{} exhibits different trends at low and high $t$ values and the number of training data samples reduces with increasing both $t$ and $\rxpowerThreshold$.
Since the \MB{} approach is highly biased to the \wbl{} model, the accuracy of its mean and variance estimations is high only in the regimes where the majority of the training data lies, and degrades with increasing $t$ and $\rxpowerThreshold$ as illustrated in Figs. \ref{fig:meanChange} and \ref{fig:varChange}.
In contrast, due to having lower bias, the \DD{} approach generalizes throughout all $t$ and $\rxpowerThreshold$, but with a price of significant accuracy losses in the mean and variance estimations.
\section{Conclusions}\label{sec:conclusion}

In this letter, we have analyzed the non-blocking connectivity of \gls{urc} systems through the lens of model-based and data-driven methods in order to estimate connectivity statistics using a set of non-blocking connectivity duration training samples.
Therein, we have measured the reliability of the connectivity by using statistical tools from survival analysis.
We have also validated our analysis based on simulations.
The results show that the \wbl{} model-based method can be accurately estimated with low sample complexity and characterizes well the tail events without the knowledge on the channel statistics.
In contrast, the data-driven design aligns well with the highly probable events under large sizes of training data highlighting the bias-variance tradeoff between the aforementioned two approaches.
Finally, this work provides insights about the choice of transmit power in terms of channel blocking statistics.
Future work will investigate hybrid approaches combining both data-driven and model-driven techniques.

\appendices
\section{Proof of Proposition \ref{thm:wbl_moments}}\label{appndx:wbl_moments}

Let $T = t + \losDuration$.
By differentiating \eqref{eqn:fail_rate_wbl}, the conditional \gls{pdf} is found as 
$f_{t}(T) = \frac{\distributionShape}{\distributionScale^\distributionShape} T^{\distributionShape-1} e^{-(T/\distributionScale)^\distributionShape}e^{(t/\distributionScale)^\distributionShape}$ for all $T\geq t$. 
Then, the $N$th moment is given by $\expect [T^N] = \int_{t}^{\infty} 
\frac{\distributionShape}{\distributionScale^\distributionShape} T^{N+\distributionShape-1} e^{-(T/\distributionScale)^\distributionShape}e^{(t/\distributionScale)^\distributionShape}
dT$.
Using the change of variables with $z=(T/\distributionScale)^\distributionShape$ and $dT = \distributionScale z^{1/\distributionShape -1} dz$,
\begin{align*}
	\expect [T^N] 
	& \textstyle
	= \int_{(t/\distributionScale)^\distributionShape}^{\infty} 
	\distributionScale^N z^{N/\distributionShape} e^{-z}e^{(t/\distributionScale)^\distributionShape}
	dz, \\
	&= \distributionScale^N e^{(t/\distributionScale)^\distributionShape}
	\Gamma\big( (t/\distributionScale)^\distributionShape; 1 + N/\distributionShape \big),
\end{align*}
where $\Gamma(\alpha,\beta) = \int_{\alpha}^{\infty} x^{\beta-1} e^{-x} dx$ is the \emph{upper incomplete gamma} function.

\bibliographystyle{IEEEtran}
\bibliography{IEEEabrv,mybib_urllc_letter}

\end{document}